\documentclass{article}
\usepackage[T1]{fontenc}
\usepackage{times}
\usepackage{a4}
\usepackage{graphics}
\usepackage{setspace}
\makeatletter

\newcommand{\LyX}{L\kern-.1667em\lower.25em\hbox{Y}\kern-.125emX\@}

\makeatother

\begin{document}

\title{Nuclear Transparency and Effective NN Cross Section in Heavy Ion Collisions
at 14.6 GeV/nucleon}

\author{H.M. Ding\thanks{
On leave of absence from Department of Physics, Jilin University, Changchun,
China.
}, J. Hüfner\thanks{
Corresponding Address: Institut für Theoretische Physik, Universität Heidelberg,
Philosophenweg 19, D-69120 Heidelberg, Germany. Tel: 06221-549440; Fax: 06221-549331;
Email: Joerg.Huefner@urz.uni-heidelberg.de.
} \emph{}\\
\emph{Institut für Theoretische Physik, Universität Heidelberg,} \\
\emph{Philosophenweg 19, D-69120 Heidelberg, Germany}}

\maketitle
\begin{abstract}
The probability of a projectile nucleon to traverse a target nucleus without
interaction is calculated for central \( Si \)-\( Cu \) and \( Si \)-\( Pb \)
collisions. Special attention is given to the impact parameter range which contributes
to events with large transverse energy. A fit to the data from E814 requires
an effective NN cross section of \( \sigma _{eff}=54.2\pm 5.0 \) mb (compared
to the free space value \( \sigma _{in}^{NN}=30 \) mb ) and is interpreted
as one related to wounded target nucleons.


\end{abstract}
The dynamics of a heavy ion collision is a complicated many-body problem for
various reasons. It is the task of appropriately designed experiments to isolate
one particular aspect of the dynamics and elucidate its physics. Wounded nucleons
are one of the open problems. In a heavy ion collision a nucleon may undergo
a sequence of collisions, which follow so rapidly one after another, that the
nucleon is no more in its ground state, even not necessarily in any definite
excited baryonic resonance ( like \( \Delta  \) or \( N^{*} \) ). We will
speak of a wounded nucleon. In a next encounter with another nucleon this wounded
nucleon will not interact with the free space NN cross section \( \sigma _{in}^{NN} \)
but with an effective one \( \sigma _{eff} \) . Is it possible to determine
\( \sigma _{eff} \) from experiment?

The E814 collaboration has designed an experiment to answer this question\cite{1}.
At the energy of 14.6 GeV/nucleon, the projectile \( ^{28}Si \) collides with
\( Al \), \( Cu \) and \( Pb \) and the beam rapidity nucleons are studied
as a function of the centrality of the reaction (controlled by a measurement
of the transverse energy ). The beam-rapidity nucleons belong to the projectile
and have not lost any energy in the reaction. Of course, in a peripheral reaction
one always has the so called spectator nucleons, which pass the target nucleus
without interaction. They are not interesting for our purpose. However , in
a central event, e.g., for \( Si \) on \( Pb \), one still sees beam rapidity
nucleons. The target nucleus is transparent for these projectile nucleons. We
expect the transparency of a heavy nucleus like \( Pb \) to be small. Indeed,
the ``survival probability'' \( S \) for a projectile nucleon to pass through
the target without any inelastic interaction has been measured to \( S_{exp}=3.5\cdot 10^{-3} \)
for central \( Si \)-\( Pb \) collisions\cite{1}. Does this result contain
information about wounded nucleons and effective NN cross sections? 

In a heavy ion collision those projectile nucleons which arrive first at the
target nucleus, interact with ground state nucleons of the target, but may transform
them into wounded ones. The next wave of projectile nucleons already finds wounded
target nucleons and interacts with them via a modified, probably larger cross
section and transforms them to a higher degree of woundedness. This procedure
repeats itself for the third and following waves. A complicated situation like
this may best be simulated by a cascade code. However, present day computer
codes are not yet in a position to handle off-energy-shell situations. Instead
they use on-shell baryons ( in their ground and excited states ) together with
mesons as their basic degrees of freedom. A code like this, ARC, has been applied
to the data from E814 by Schlagel et al. \cite{2} and has lead to agreement
with the data without wounded nucleons and effective cross section, though with
large error bars because of the low statistics. As we will argue on the basis
of time scales, the basic assumptions of the cascade code may not be fulfilled
and another approach may be better justified. It will be presented in this paper.

The survival probability \( S \) for a projectile nucleon to pass through the
target without inelastic interaction is calculated in a Glauber type approximation
(straight lines, frozen nucleons) as 

\begin{eqnarray}
S(b,\sigma _{eff})= & \int d^{2}sT_{p}(\vec{b}-\vec{s})e^{-\sigma _{eff}A_{t}T_{t}(\vec{s})},\label{prob15} 
\end{eqnarray}
 where \( \vec{b} \) is the impact parameter of the nucleus-nucleus collision,
\( T_{p}(b) \) and \( T_{t}(b) \) are the thickness functions (\( T(b)=\int dz\rho (b,z), \)
\( \int d^{3}x\rho (x)=1 \)) of projectile and target, respectively. The straight
line geometry is certainly justified for the through-going nucleons. The use
of ``frozen'' nucleons and the neglect of any other degrees of freedom, like
mesons, needs justification. We consider a target nucleon and estimate the time
\( \Delta t \) in its rest system which has elapsed between the arrival times
of the first and the last projectile nucleons: \( \Delta t\leq 2R_{p}/\gamma _{p} \)
where \( 2R_{p}\simeq 7fm \) is the diameter of the \( Si \) projectile and
\( \gamma _{p}\simeq 15.6 \) is the Lorenz factor for the experiment under
consideration. The time \( \Delta t\approx 0.5fm/c \) is rather short: (i)
The target nucleon has not moved significantly in space ( frozen approximation
is good ). (ii) According to the uncertainty principle the intrinsic excitation
energy is a uncertain by \( \Delta E\geq (\Delta t)^{-1}\approx 0.4 \)GeV and
does not allow the definition of a definite excited state. (iii) Secondary hadrons
are not yet formed since typical formation times are of order 1 fm/c. 

The nature of a wounded nucleon is not clear. We parametrise any modification
into an effective cross section \( \sigma _{eff} \) between a ground-state
nucleon (of the projectile) and a wounded nucleon (of the target), and determine
it from experiment. Using Saxon-Woods parametrisations for the densities \( \rho _{t} \)
and \( \rho _{p} \) with the surface thickness \( a=0.52 \) fm for all nuclei
and the half-density radius \( r_{A} \) such that the root-mean-square radius
of the nucleus equals the charge radius\cite{3}, the survival probability \( S(b,\sigma _{eff}) \)
is calculated for two values \( \sigma _{eff}=30 \) and 50mb, and the results
are displayed in Fig.1. The experimental point, measured at transverse energy
\( E_{t}^{c}=15.5 \) GeV is also shown in the figure. It is obtained from the
measured mean multiplicity \( <M_{c}> \) of beam rapidity protons by

\begin{equation}
S(E_{t})=<M_{c}>(E_{t})/Z,
\end{equation}
where \( Z=14 \) is the number of protons in \( Si \). We have assumed - as
the authors of the experiment do - that the high value \( E^{c}_{t} \) corresponds
to a central collision which is assigned a value \( b=0 \). Then the experimental
value is close to the curve \( \sigma _{eff}=30mb \). In fact the equation
\( S(b=0,\sigma _{eff})=S_{exp}(E_{t}^{c}=15.5GeV) \) leads to a value \( \sigma _{eff}=31.1\pm 0.7 \)
mb. This result reproduces a similar calculation using uniform density distributions
by the E814 collaboration \cite{1}, who have concluded that \( \sigma _{eff}=\sigma _{in}^{NN} \)
within error bars and no anomaly being visible. 

The crucial step in the argument is the assignment of \( b=0 \) to the central
value of transverse energy \( E_{t} \). Indeed , a given value of \( E_{t} \)
determines an impact parameter \( b_{m}(E_{t}) \) only within a band \( \Delta b(E_{t}) \).
And \( \Delta b(E_{t}) \) may be very large! We study the relation between
\( E_{t} \), \( b_{m} \) and \( \Delta b \) in the form of a probability
distribution \( P(E_{t},b) \), which gives the probability that values of \( b \)
contribute to events with a given value of \( E_{t} \). We normalize it as
\( \int dE_{t}P(E_{t},b)=1 \). With this function and the differential inelastic
heavy ion cross section \( d\sigma _{in}/d^{2}b \), one can obtain the dependence
of the survival function \( S(E_{t},\sigma _{eff}) \) as function of transverse
energy

\begin{equation}
\label{pet}
S(E_{t},\sigma _{eff})=\frac{\int d^{2}bS(b,\sigma _{eff})P(E_{t},b)d\sigma _{in}/d^{2}b}{d\sigma _{in}/dE_{t}},
\end{equation}
 where

\begin{equation}
\label{s_et}
\frac{d\sigma _{in}}{dE_{t}}=\int d^{2}bP(E_{t},b)\frac{d\sigma _{in}}{d^{2}b}.
\end{equation}
We will use Eq.(\ref{s_et}) to determine \( P(E_{t},b) \) from a comparison
with the measured transverse energy distribution \( d\sigma _{in}/dE_{t} \).
Only then \( S(E_{t},\sigma _{eff}) \) can be calculated from Eqs.(\ref{prob15})
and (\ref{pet}) without ambiguity. We discuss our parametrisations.
\vspace{0.3cm}

The inelastic cross section \( d\sigma _{in}/d^{2}b \) is taken from the folding
model

\begin{equation}
\label{s_b}
\frac{d\sigma _{in}}{d^{2}b}(b)=1-exp[-\sigma ^{NN}_{in}A_{p}A_{t}\int d^{2}sT_{p}(\vec{b}-\vec{s})T_{t}(s)].
\end{equation}
For central collisions (small value of b ), the exponential is practically zero
and any uncertainties, e.g., the choice of \( \sigma _{in}^{NN} \), are unimportant.
The parametrisation of \( P(E_{t},b) \) is more model dependent. We choose
a Gaussian parametrisation

\begin{equation}
\label{p_et_b}
P(E_{t},b)=\frac{1}{\sqrt{2\pi \sigma ^{2}_{t}(b)}}exp\{-\frac{[E_{t}-E_{t}(b)]^{2}}{2\sigma ^{2}_{t}(b)}\},
\end{equation}
which satisfies the normalization condition. We make the usual assumptions\cite{4}\cite{5}
for the functions \( E_{t}(b) \) and \( \sigma _{t}(b) \)

\begin{equation}
E_{t}(b)=N(b)\epsilon _{0},
\end{equation}

\begin{equation}
\label{st_b}
\sigma ^{2}_{t}(b)=N(b)\epsilon _{0}^{2}\omega .
\end{equation}
 Here \( N(b) \) is calculated either in the `` collision model'' \( N_{c}(b) \)\cite{4}
or in the ``wounded nucleon model'' \( N_{w}(b) \)\cite{5},

\begin{equation}
N_{c}(b)=\sigma _{in}^{NN}A_{t}A_{p}\int d^{2}bT_{p}(\vec{s})T_{t}(\vec{b}-\vec{s}),
\end{equation}

\begin{equation}
N_{w}(b)=A_{t}\int d^{2}bT_{p}(s)exp[-\sigma _{in}^{NN}T_{t}(\vec{b}-\vec{s})]+(t\longleftrightarrow p).
\end{equation}
 \( N_{c}(b) \) equals the mean number of NN collisions in a projectile-target
interaction with impact parameter \( b \), and \( N_{w}(b) \) equals the number
of nucleons in the overlap volume of projectile and target. The proportionality
constant \( \epsilon _{0} \) between the observed transverse energy \( E_{t} \)
and \( N(b) \) depends on the dynamics of hadron production, but also on the
experimental set up (chosen rapidity interval and acceptance of counter). It
will be a fit parameter. In the collision model, which we will use, only the
product \( \epsilon _{0}\sigma _{in}^{NN} \) enters and the physics of Eq.(\ref{p_et_b})
is completely independent of the choice of \( \sigma _{in}^{NN} \). For the
wounded nucleon model this is only approximately true. The ansatz \( \sigma ^{2}_{t}\propto N(b) \)
corresponds to the hypothesis that we deal with statistical fluctuations in
\( N(b) \). It is known, however, that the proportionality constant \( \omega  \)
in Eq.(\ref{st_b}) depends strongly on the rapidity interval of the accepted
particles\cite{4}. The reason is not clear\cite{6}. We take \( \omega  \)
as a free parameter. 

Using expressions (\ref{s_b}) to (\ref{st_b}) we have calculated \( d\sigma _{in}/dE_{t} \)
and have varied the parameters \( \epsilon _{0} \) and \( \omega  \) until
the data for \( Si \)-\( Pb \) are fitted. The result is shown in Fig.2 with
\( \sigma _{in}^{NN}=30mb \), \( \epsilon _{0}=0.067GeV \), \( \omega =7.8 \)
for the collision model and \( \epsilon _{0}=0.1GeV,\omega =6. \) for the wounded
nucleon model. Both models for the calculation of \( N(b) \), describe the
data \( Si \)-\( Pb \) equally well. But when the same parameters are used
to describe \( d\sigma _{in}/dE_{t} \) for \( Si \)-\( Al \) and \( Si \)-\( Cu \)
collisions, the collision model gives better fits and therefore we discard the
other model. 
\vspace{0.3cm}

According to Eq.(\ref{pet}), \( P(E_{t},b) \) determines the integration region
in impact parameter \( b \), which contributes to the integral for a given
value of \( E_{t} \). As an example, Fig.1 shows the weight function \( P(E^{c}_{t},b) \).
Its width is unexpectedly large. In particular the survival probability \( S(b,\sigma _{eff}) \)
changes considerably in the \( b \)-region, where \( P(E^{c}_{t},b) \) is
large. This invalidates the assumption to put the experimental point at \( b=0 \).

Since \( P(E_{t},b) \) plays such important role in the understanding of the
experiment, we have studied its properties in more detail. For the functions
\( P(E_{t},b) \) used to fit \( d\sigma _{in}/dE_{t} \) we have calculated
the value \( b_{m}(E_{t}) \) of the impact parameter which contributes maximally
to a given \( E_{t} \) and \( \Delta b(E_{t}) \) which is the width of the
band of impact parameters around \( b_{m}(E_{t}) \). We define these two quantities
by 

\begin{equation}
\label{mb}
\frac{d}{db}P(E_{t},b)\mid _{b=b_{m}(E_{t})}=0,
\end{equation}
and

\begin{equation}
\label{mb2}
[\Delta b(E_{t})]^{2}=\int d^{2}b(b-b_{m})^{2}P(E_{t},b)/\int d^{2}bP(E_{t},b).
\end{equation}
For the Gaussian parametrisation Eq.(\ref{p_et_b}) one has approximately the
relations 

\begin{equation}
E_{t}[b_{m}(E_{t})]=E_{t},
\end{equation}

\begin{equation}
[\Delta b(E_{t})]^{2}=\omega \frac{N(b_{m})}{[N'(b_{m})]^{2}},
\end{equation}
where \( N'(b)=dN/db \). Note that the width of the band in impact parameter
, which contributes to a given \( E_{t} \) is proportional to the width \( \omega  \)
of the Gaussian of \( P(E_{t},b) \). 
\vspace{0.3cm}

Fig.3 shows the position \( b_{m}(E_{t}) \) of the maximum of \( P(E_{t},b) \)
and the width \( \Delta b(E_{t}) \) as a function of \( E_{t} \) for \( Si \)-\( Pb \)
collisions. As expected the position \( b_{m} \) goes to zero and then stays
zero for large values of \( E_{t} \). However, the width \( \Delta b \) shows
a break at \( E_{t}^{0} \) where \( b_{m}(E_{t}) \) becomes zero. This is
related to a sudden change in the shape of \( P(E_{t},b) \) at \( E_{t}^{0} \).
For \( E_{t}>E_{t}^{0} \) the width \( \Delta b \) decreases only very slowly.

Fig.1 shows as example the distribution \( P(E_{t},b) \) for a central collision
( \( E_{t}^{c}=15.5 \) GeV ). We have compared this form with the corresponding
one \( P_{U}(E_{t}^{c},b) \) calculated \cite{8} with the help of the code
\( UrQMD \)\cite{9}. The shape of \( P_{U} \) is similar to \( P \), but
narrower. (The value of \( <b> \) is smaller by about 25\%). However, also
\( d\sigma /dE_{t} \) calculated from \( UrQMD \) falls off faster than the
experimental data. Using the functions \( S(b,\sigma _{eff}) \) and \( P(E_{t}^{c},b) \)
as shown in Fig.1, the calculated value of \( S(E_{t}^{c},\sigma _{eff}) \)
gets contributions from a considerable range of values of \( b \). We find 

\begin{equation}
S(E_{t}^{c},\sigma _{eff})/S(b=0,\sigma _{eff})=2.86
\end{equation}
for \( \sigma _{eff}=30mb \). Since \( S(b=0,\sigma _{eff}=30mb) \) corresponded
essentially to the experimental value, \( S(E_{t}^{c},\sigma _{eff}=30mb) \)
does not. A solution of the equation for \( \sigma _{eff} \)

\begin{equation}
S(E_{t}^{c},\sigma _{eff})=S_{exp}(E_{t}^{c})
\end{equation}
leads to 

\[
\sigma _{eff}=54.2\pm 5.0\pm 6.4\pm 5.6mb.\]
 The meanings of the errors are explained below and in the caption of Table
1. The same equation for central events ( \( E_{t}^{c}=9.5 \) GeV )in \( Si \)-\( Cu \)
collisions leads to \( \sigma _{eff}=50.3\pm 15.0\pm 4.6\pm 6.4mb \) , in agreement
with the value from \( Si \)-\( Pb \). Since \( Al \) is a smaller nucleus
than \( Si \), the beam rapidity nucleons from \( Si \)-\( Al \) will always
be contaminated by spectator nucleons and not very sensitive to effects of the
nuclear transparency. The results of the values for \( \sigma _{eff} \) are
compiled in Table 1. We have also calculated \( \sigma _{eff} \) by using \( P_{U}(E_{t}^{c},b) \)
from \( UrQMD \) and have obtained \( \sigma _{eff}=40.3\pm 3.0mb \), where
the error reflects only the statistical fluctuations in \( P_{U} \). 

The result for \( \sigma _{eff} \) from our analysis is at variance with the
conclusion by the authors of the E814 experiment \cite{1} and of the cascade
calculation of Ref. \cite{2}. While we are unable to understand the difference
to the cascade calculation\cite{2}, the difference to the argument of the E814
collaboration is clear: They assume that central collisions means \( b=0 \)
with very small width, while our analysis shows that even for central collisions
a fairly wide band of impact parameters contributes. The width of this band
depends on the parameters \( \epsilon _{0} \) and \( \omega  \) in the function
\( P(E_{t},b) \). How sensitive are the data to these parameters? Table 1 gives
the uncertainties in the extracted values of \( \sigma _{eff} \) if one assumes
uncertainties of 5\% in \( \epsilon _{0} \) and 10\% in \( \omega  \), respectively.
A reduction by 25\% in \( \omega  \) would reproduce the results of \( UrQMD \)
for \( d\sigma /dE_{t} \) and \( \sigma _{eff} \) but would be in contradiction
to the data for \( d\sigma /dE_{t} \) and therefore the extracted value \( \sigma _{eff}\approx 40mb \)
is doubtful. According to the results presented in Table 1 and from the cascade
calculations the extracted effective cross section seems to be definitely larger
than the inelastic cross section \( \sigma _{in}^{NN} \) for a NN collision
in vacuum. 

The difference may be interpreted in terms of wounded nucleons. We try to estimate
the effect by the following model\cite{10} which a wounded nucleon \( N^{+} \)
consists of a nucleon plus a semi-hard prompt gluon, which is radiated as bremsstrahlung
in quark-quark interactions during an inelastic NN collision\cite{11}. We furthermore
use the additive quark model ( a nucleon consists of 3 quarks ) to relate the
\( NN \) cross section to the nucleon-quark (\( Nq \)) one via \( \sigma ^{NN}=3\sigma ^{Nq} \).
Then

\begin{equation}
\sigma ^{NN^{+}}=3\sigma ^{Nq}+\sigma ^{Ng}=\sigma ^{NN}(1+\frac{\sigma ^{Ng}}{3\sigma ^{Nq}}).
\end{equation}
If in addition one assumes \( \sigma ^{Ng}/\sigma ^{Nq}\approx 9/4 \), the
color factor, one obtains \( \sigma _{in}^{NN^{+}}\approx 50mb \). In view
of all the uncertainties in the definition of \( \sigma _{eff} \), in the model
for \( N^{+} \) and in the additive quark model, the close agreement between
experiment and the theoretical estimate must be considered fortuitous, but it
may show that the order of magnitude of the extracted value for the effective
cross section is not unreasonable.

Acknowledgement: We thank P. Braun-Munzinger, S. Kahana, B. Kopeliovich and
J. Stachel for several discussions and W. Cassing and P. Glässel for running
cascade codes for us. H.M. Ding is grateful to the Institut für Theoretische
Physik Universität Heidelberg for its hospitality and is grateful for a fellowship
of the government of the People's Republic of China. The research is partly
supported by the BMFT under grant O6HD856.

\textbf{\( \qquad  \)}

\textbf{\( \qquad  \)}

\textbf{\( \qquad  \)}

\textbf{Captions:}

\textbf{Fig.1} Survival probabilities \( S(b,\sigma _{eff}) \) of a beam rapidity
nucleon calculated for \( \sigma _{eff}=30 \) mb and \( \sigma _{eff}=50 \)
mb (solid lines) and the correlation \( P(E^{c}_{T}=15.5GeV,b) \) between transverse
energy and impact parameter (dashed line) for \( Si \)-\( Pb \) collisions
as a function of \( b \). The data point is obtained from the measured multiplicity
\( <M_{c}>=0.049\pm 0.005 \) of beam rapidity protons at \( E_{t}^{c} \) .

\textbf{Fig.2} Transverse energy distributions in the rapidity range \( -0.5<\eta <0.8 \)
for a \( Si \) beam at 14.6GeV/nucleon on different targets. Data are taken
from {[}7{]}. The calculations refer to the collision model (solid lines) and
the wounded nucleon model (dashed curves). For each model one set of parameters
\( \epsilon _{0} \), \( \omega  \) has been determined by a fit to the \( Pb \)
data and then applied to the other data.

\textbf{Fig.3} The position \( b_{m}(E_{t}) \) of the maximum of \( P(E_{t},b) \)
in impact parameter \( b \) (solid line ) and the width \( \Delta b(E_{t}) \)
around it (dashed line ) as a function of \( E_{t} \) for \( Si \)-\( Pb \)
collisions.

\textbf{Table 1} The fitted effective NN cross sections from the experimental
value of the multiplicity \( <M_{c}> \) of beam rapidity nucleons in collisions
of \( Si \)-\( Pb \) and \( Si \)-\( Cu \) at AGS energy of 14.6GeV/nucleon.
If one assumes that the \( E_{t}^{c} \) corresponds to \( b=0 \) and uses
distributions of uniform density (u.d.) or Saxon-Woods (S.W.) one finds the
values in columns 4 {[}1{]} and 5, respectively. The last column corresponds
to our result. The first error corresponds to the uncertainty in \( <M_{c}> \),
while the second and third errors correspond to uncertainties of 5\% in \( \epsilon _{0} \)
and 10\% in \( \omega  \), respectively. 

\textbf{\( \qquad  \)}

\textbf{\( \qquad  \)}

\textbf{\( \qquad  \)}

\textbf{\( \qquad  \)}

\textbf{\( \qquad  \)}

\textbf{\( \qquad  \)}

\textbf{\( \qquad  \)}

\textbf{\( \qquad  \)}

\textbf{\( \qquad  \)}

\textbf{\( \qquad  \)}

\textbf{\( \qquad  \)}

\vspace{0.3cm}
{\par\centering \resizebox*{1\textwidth}{0.5\textheight}{\rotatebox{270}{\includegraphics{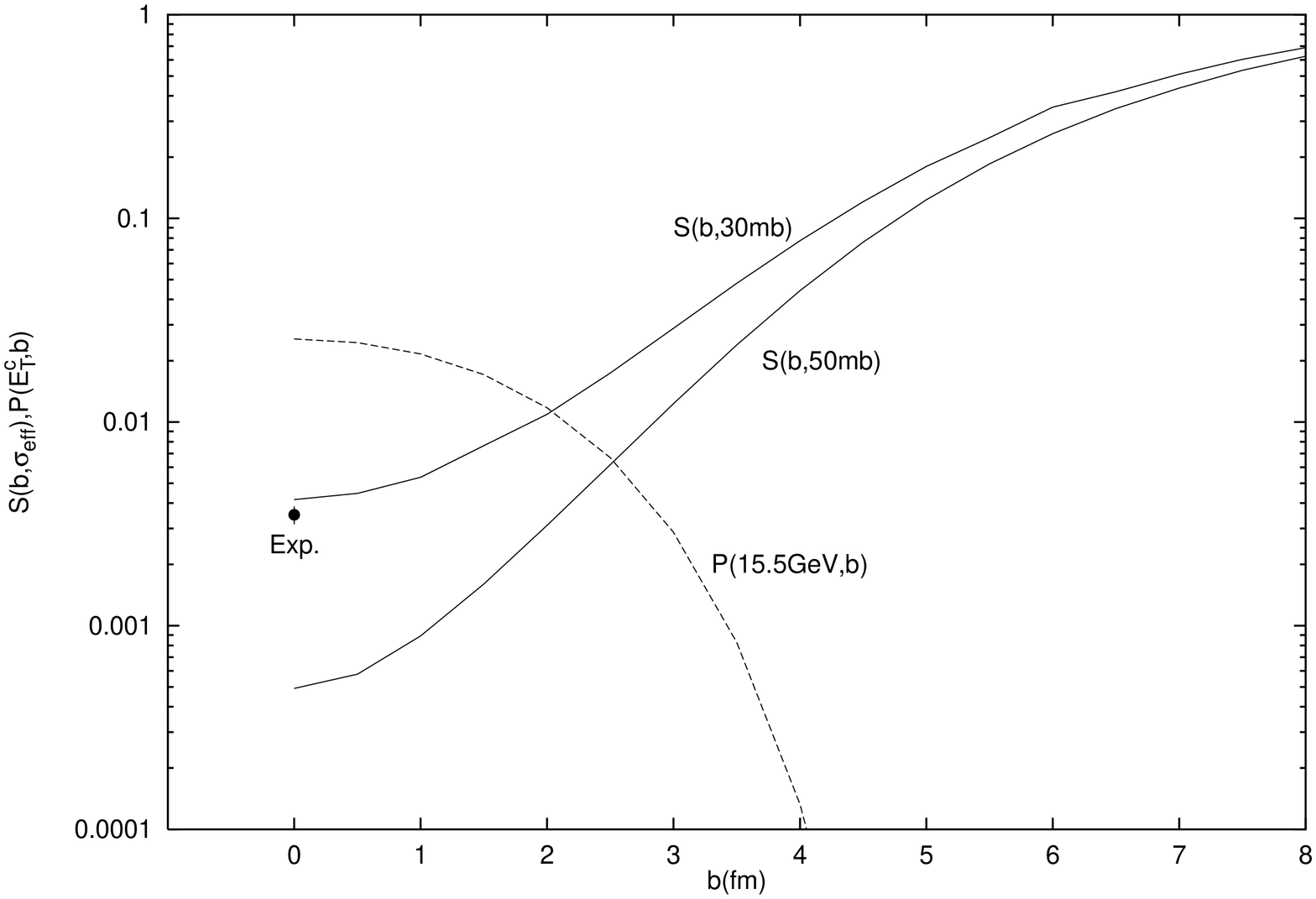}}} \par}
\vspace{0.3cm}

{\par\centering \textbf{Fig.1 }\par}
\vspace{0.3cm}

\textbf{\( \qquad  \)}

\textbf{\( \qquad  \)}

\textbf{\( \qquad  \)}

\textbf{\( \qquad  \)}

\textbf{\( \qquad  \)}

\textbf{\( \qquad  \)}

\textbf{\( \qquad  \)}

\textbf{\( \qquad  \)}

\textbf{\( \qquad  \)}

\textbf{\( \qquad  \)}

\textbf{\( \qquad  \)}

\textbf{\( \qquad  \)}

\textbf{\( \qquad  \)}

\vspace{0.3cm}
{\par\centering \resizebox*{1\textwidth}{0.5\textheight}{\rotatebox{270}{\includegraphics{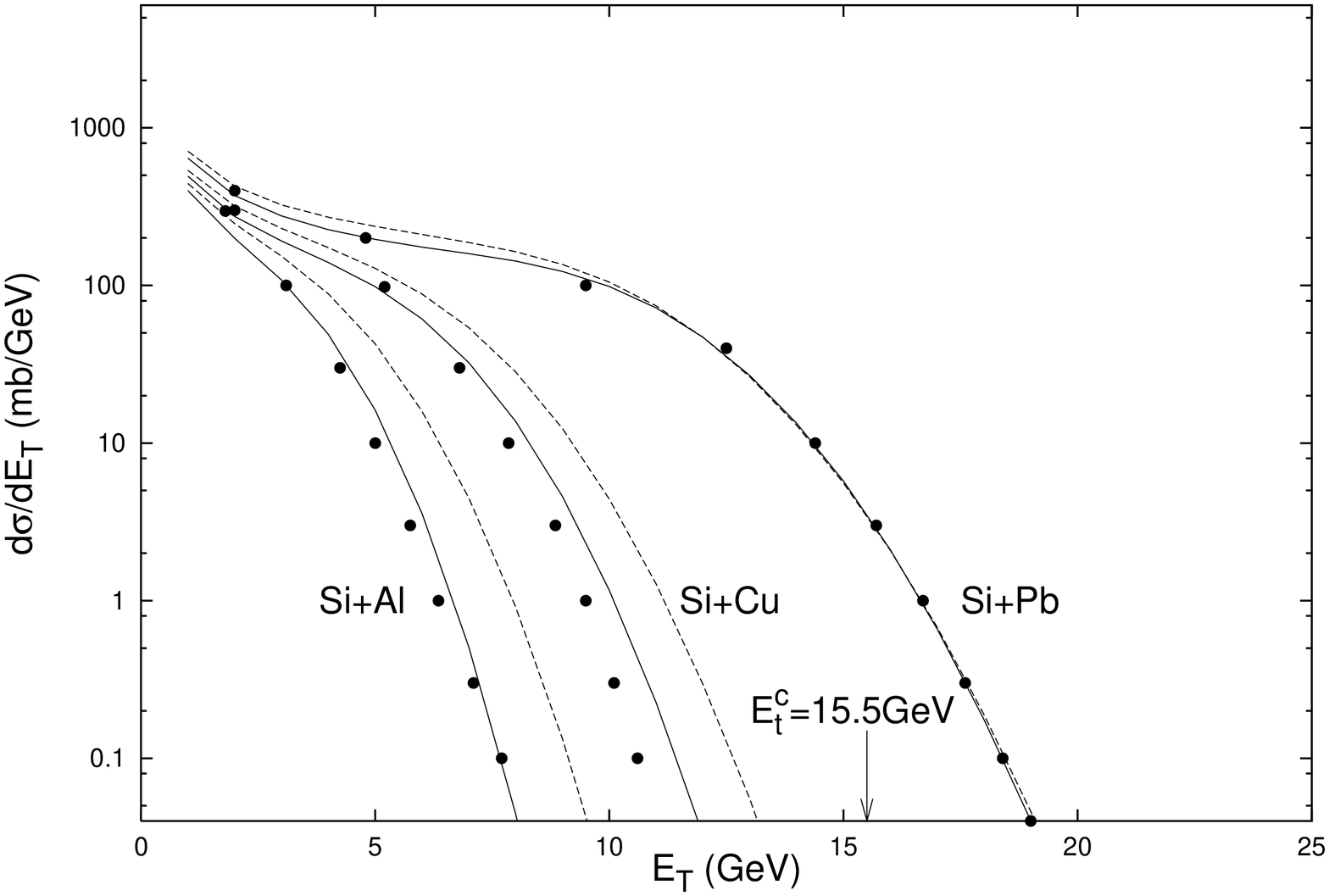}}} \par}
\vspace{0.3cm}

{\par\centering \textbf{Fig.2} \par}

\textbf{\( \qquad  \)}

\textbf{\( \qquad  \)}

\textbf{\( \qquad  \)}

\textbf{\( \qquad  \)}

\textbf{\( \qquad  \)}

\textbf{\( \qquad  \)}

\textbf{\( \qquad  \)}

\textbf{\( \qquad  \)}

\textbf{\( \qquad  \)}

\textbf{\( \qquad  \)}

\vspace{0.3cm}
{\par\centering \resizebox*{1\textwidth}{0.5\textheight}{\rotatebox{270}{\includegraphics{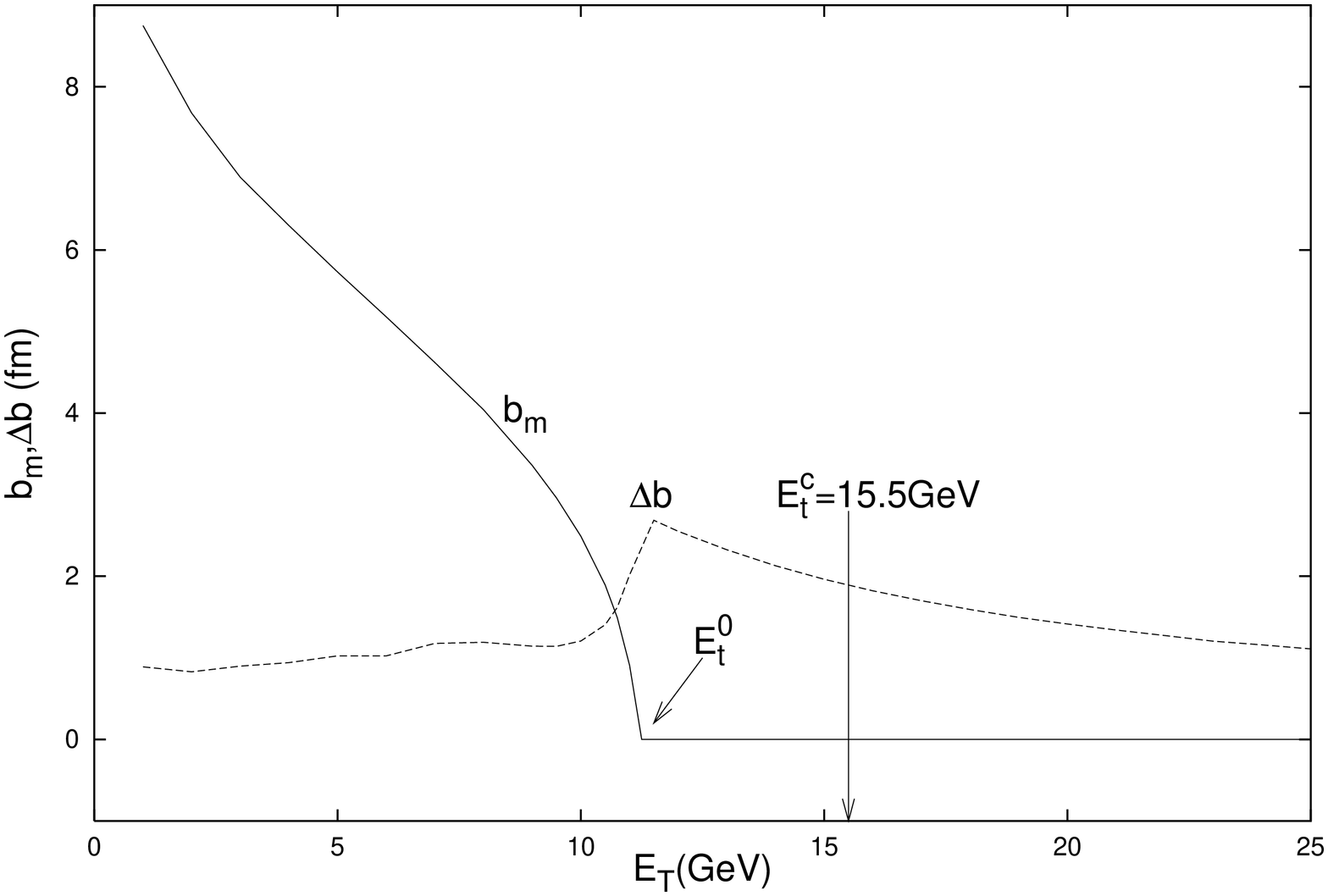}}} \par}

{\par\centering \textbf{Fig.3} \par}
\vspace{0.3cm}

\textbf{\( \qquad  \)}

\textbf{\( \qquad  \)}

\textbf{\( \qquad  \)}

\textbf{\( \qquad  \)}

\textbf{\( \qquad  \)}

\textbf{\( \qquad  \)} \textbf{\( \qquad  \)}

\textbf{\( \qquad  \)}

\textbf{\( \qquad  \)}

\textbf{\( \qquad  \)}

\textbf{\( \qquad  \)} \textbf{\( \qquad  \)}

\textbf{\( \qquad  \)}

\textbf{\( \qquad  \)}

\textbf{\( \qquad  \)}

\textbf{\( \qquad  \)}

\textbf{\( \qquad  \)}

\textbf{\( \qquad  \)}

\textbf{\( \qquad  \)}

\textbf{\( \qquad  \)}

\begin{table}
\vspace{0.3cm}
{\centering \begin{tabular}{|c|c|c|ccc|}
\hline 
&
\multicolumn{1}{|c|}{}&
&
&
\( \sigma _{eff} \)(mb) &
 \\
\cline{4-4} \cline{5-5} \cline{6-6} 
&
\multicolumn{1}{|c|}{\( E^{c}_{T} \)(GeV)}&
\( <M_{c}> \)&
\multicolumn{1}{|c}{b=0(u.d.)}&
\multicolumn{1}{c}{b=0(S.W.)}&
\multicolumn{1}{c|}{\( E_{t}=E_{t}^{c} \)}\\
\hline 
Pb&
15.5&
0.049\( \pm  \)0.005&
28.8\( \pm  \)0.5&
31.1\( \pm  \)0.7&
54.2\( \pm  \)5.0\( \pm  \)6.4\( \pm  \)5.6\\
\hline 
Cu&
9.5&
0.66\( \pm  \)0.09&
28.8\( \pm  \)1.8&
35.8\( \pm  \)2.5&
50.3\( \pm  \)15.0\( \pm  \)4.6\( \pm  \)6.4\\
\hline 
\end{tabular}\par}\vspace{0.3cm}

\textbf{Table 1}
\end{table}

\end{document}